\documentclass[10pt,letterpaper]{article} 
\usepackage[english]{babel}
\newcommand{\lb}{\langle}
\newcommand{\rb}{\rangle}

\newcommand{\beq}{\begin{equation}}
\newcommand{\eeq}{\end{equation}}
\newcommand{\lbl}{\label}
\newcommand{\beqnar}{\begin{eqnarray}}
\newcommand{\eeqnar}{\end{eqnarray}}
\newcommand{\beqnars}{\begin{eqnarray*}}
\newcommand{\eeqnars}{\end{eqnarray*}}

\newcommand{\goesto}{\rightarrow}
\newcommand{\s}{\\[1ex]}
\newcommand{\re}[1]{(\ref{#1})}
\newcommand{\q}{\quad}

\newcommand{\tnr}{\otimes}

\newcommand{\trM}{\mbox{tr}_M\,}

\newtheorem{definition}{Definition}
\newcommand{\bR}{{\bf R}} 
\newcommand{\NE}{N\!E}
\begin{document}
\begin{center}
\Large
\bf
Quantum weak values are not unique \\
\large
What do they actually measure?  
\\[2ex] 
\rm
Stephen Parrott% 
\footnote{For contact information, go to 
www.math.umb.edu/$\sim$sp.  
}
\\[3ex]
\end{center}
\begin{abstract}
Precise definitions of ``weak [quantum] measurements'' and 
``weak value'' [of a quantum observable] are offered, 
which seem to capture the meaning of the often vague ways 
that these terms are 
used in the literature.
Simple finite dimensional examples are given showing that
weak values of an observable are not unique, and in fact arbitrary 
weak values can be obtained by appropriate weak measurements.  
This implies that a ``weak value'' of an observable $A$, {\em by itself}, 
can furnish no unambiguous information about $A$; any information in a weak value
is inextricably connected with the particular measurement procedure 
used to obtain that weak value.   Moreover, 
arbitrary weak values can be obtained
using a ``meter space'' of dimension as small as 2.

A ``Remarks'' section questions the utility of ``weak measurement''. 
\end{abstract} 
\section{Introduction and notation}
\label{sec: intro}
We assume that the reader is familiar with the concept of ``weak value''
of a quantum observable.  This concept was introduced in the seminal paper
\cite{AAV} of Aharonov, Albert, and Vaidman, called  ``AAV'' below.  It 
will be briefly reviewed below, and
a much fuller presentation intended for those  unfamiliar with weak values 
can be found in \cite{parrott}. 

	We attempt to stay as close as possible to traditional physics notation,
	reverting to notation more common in mathematics only when it seems 
	less ambiguous or complicated. 
	Our mathematical formulation of quantum mechanics 
generally follows that of 
	Chapter 2 of the  book of Nielsen and Chuang \cite{N/C}, 
with differences in notation noted below.  

	The inner product of vectors $v, w$ in a complex Hilbert space $H$  will be
	denoted $\lb v, w \rb$, with the physics convention that this be linear in
	the second variable $w$, and conjugate-linear in the first variable $v$.
	The norm of a vector $v$ will be denoted as 
	$ |v| := \lb v,v \rb ^{1/2}$.  

Technically, a (pure) ``state'' of a quantum system with Hilbert space $H$
is an equivalence class of nonzero 
vectors in $H$, where vectors $v, w \in H$ are
equivalent if and only if $w = \alpha v$ for some nonzero constant $\alpha$.
However, we informally refer to vectors in $H$ as
``states'', or ``pure states'', when we need to distinguish 
between pure states and ``mixed states'' (see below).  
A state $v$ is said to be {\em normalized} if $|v| = 1$.   
We do not assume that states are necessarily normalized.

	The projector to a subspace $E$ will  be denoted $P_E$, in 
	place of the common but unnecessarily complicated 
	physics notation $\sum_i | e_i  \rb \lb e_i |$, where $ \{e_i\}$ is an 
	orthonormal basis for $E$.  
	When $E$ is the entire Hilbert space of states, $P_E$ is called the 
	{\em identity operator} and denoted $I := P_E$.   
	When $E$ is the one-dimensional subspace
	spanned by a vector $w$, we may write $P_w$ for $P_E$.  
	When  $|w| = 1$,  $P_w v = \lb w, v \rb w$, but the reader should
	keep in mind that   under our convention, $ P_w = P_{w/|w|}$, so this 
	formula for $P_w$ only applies for $|w| = 1$. 

Mixed states are represented
by ``density matrices'' $\rho : H \rightarrow H$, which are defined as 
positive operators on $H$ of trace 1.  
A pure state $h \in H$ corresponds to the density
matrix $P_h$. 

We shall be dealing with a quantum system $S$ in which we are primarily 
interested, which will be coupled to a quantum ``meter system'' $M$.  
We make no notational distinction between the physical systems $S$ and $M$
and their Hilbert spaces. 

The composite system of $S$ together with $M$
is mathematically represented by the Hilbert space tensor product 
$S \tnr M$.  We assume the reader is generally familiar with the 
mathematical definition of $S \tnr M$.  The highlights of the definition
are as follows.

Some, but not all, vectors in $S \tnr M$ can be written in the form 
$s \tnr m$ with $s \in S$ and $m \in M$; these are called ``product states''.  
Typical physics notation for $s \tnr m$ might be 
$|s\rb |m\rb$ or 
$|s\rb_S |m\rb_M$. 
Every vector $v$ in 
$S \tnr M$ is a (possibly infinite) linear combination of product states
:  $v = \sum_i s_i \tnr m_i$.

If $\rho$ is a density matrix on $S \tnr M$, its partial trace with respect
to $M$, denoted $\trM \rho : S \rightarrow S$ is a density matrix on $S$.
The (mixed) state of $S$ corresponding to $\rho$ is $\trM \rho$.  
\section{``Weak'' measurements of a quantum observable}
\label{sec: weakm}
This section will review and formalize the notion of ``weak measurement 
protocol''.  This is preliminary to the concept of ``weak value''
of a quantum observable, which is the main subject of this paper.  

Let $S$ denote the Hilbert space of a quantum system (also denoted $S$) and
$A : S \rightarrow S$ a quantum observable (Hermitian operator) on $S$.  
To avoid technical mathematical issues,
we assume throughout that $S$ is finite dimensional, and until further
notice, we shall also assume that the meter system $M$ is finite dimensional.

A measurement of the observable $A$ when the system 
$S$ is in (pure)  state $ s  \in S $ will change (``project'') the 
pre-measurement state $s$ to one of the eigenstates of $A$. 
Thus is is impossible in general to measure $A$ without changing the state
of $S$.  However, AAV devised a clever way to measure the average value
$\lb s, As \rb$ of $A$ in the normalized state $s$ without significantly changing $s$.
This will be called a ``weak'' measurement, which will be  more precisely
defined below.
The next paragraphs describe a more general formulation of AAV's method.

Let $M$ be the Hilbert space of a ``meter'' system which will be specially
tailored below to measure $\lb s, As \rb$.  Then the Hilbert space of the 
composite system of $S$ together with $M$ is the tensor product $S \tnr M$.%

The measurement of $\lb s, As \rb$ will be accomplished by measurement in
$S \tnr M$ of an observable of the form $I \tnr B$, where $I$ denotes the
identity operator on $S$, and $B : M \rightarrow M$ is a Hermitian operator
on $M$.   We shall refer $I \tnr B$ as the ``meter observable'' and 
we think of measuring $I \tnr B$ as ``reading the meter''. 

Actually, we shall not be concerned with the detailed measurements 
of  $I \tnr B$ 
(each of which would result in one of the eigenvalues of $B$), 
but rather the average
value $\lb r, (I \tnr B) r \rb$ of $I \tnr B$ in a slightly entangled state 
$r \in S \tnr M$.%
\footnote{Because of the slight entanglement, a measurement of
$I \tnr B$ in state $r$ will give some information about a measurement
of $A \tnr I$ in state $r$.  Since we are only speaking descriptively here,
we do not need to defined ``slightly entangled''; for the curious, 
its meaning is that the state is close to a product state without actually 
being a product state. (The definition of ``entangled state'' is one which
is {\em not} a product state.)  
}
The state $r$ is obtained as follows. 

Let $m \in M$ be a given ``meter state''.  When $M$ is in state $m$ and $S$ 
in state $s$, then the composite system $S \tnr M$ is in 
the product state $s \tnr m$.  To avoid nuisance normalization factors,
assume both $s$ and $m$ are normalized:  $|s| = 1 = |m|$. 
It is traditional (though unnecessary) to apply a unitary operator to 
$s \tnr m$ to obtain the slightly entangled state $r$.%
\footnote{Some authors, such as AAV, use mathematical formulations 
which may look superficially different but are mathematically equivalent, 
as discussed 
briefly in a subsequent footnote and in detail in \cite{parrott}.
}
In the literature, the unitary
operator is generally considered to be a time evolution operator written
as $e^{-i H t}$, where $H$ is the Hamiltonian.   We shall consider
the time $t $ as a small positive parameter and emphasize this by writing
$\epsilon$ instead of $t$.  In this notation, 
\beq
\lbl{eqN5}
r = r(\epsilon) := e^{-i \epsilon H}(s \tnr m)\q, 
\eeq
with $H$ to be defined later.

Since $r(\epsilon) = e^{-i \epsilon H} (s \tnr m) = s \tnr m + O(\epsilon)$, 
and $\lb s \tnr m , (I \tnr B)(s \tnr m) \rb = \lb s , Is \rb \lb m, Bm \rb
= \lb m , Bm \rb,$ 
the average value $\lb r(\epsilon) , (I \tnr B) r(\epsilon) \rb$ of $I \tnr B$ 
in the state $r(\epsilon)$ 
will be of the form 
\beq
\lbl{eqN10}
\lb r(\epsilon) , (I \tnr B) r(\epsilon) \rb = \lb m , Bm \rb + O(\epsilon)
\q.
\eeq  
We shall assume that 
\beq
\lbl{eqN20}
\lb m , Bm \rb = 0\q.
\eeq
Physically this corresponds to the assumption that the meter initially 
reads zero, on average.  Under this assumption, 
$\lb r(\epsilon), (I \tnr B) r(\epsilon) \rb = O(\epsilon)$, so we normalize
the meter reading by dividing by $\epsilon$:
\beq
\lbl{eqN30}
\mbox{normalized average meter reading} := 
\frac{\lb r(\epsilon), (I \tnr B) r(\epsilon) \rb }
{\epsilon}
\q.
\eeq 
Eventually, we shall take a limit as $\epsilon $ goes to 0.

Our goal is to choose $B$ such that 
\beq
\lbl{eqN40}
\lim_{\epsilon \goesto 0}
 \frac{
\lb r(\epsilon), (I \tnr B) r(\epsilon) \rb }
{ \epsilon}
= \lb s, As \rb \q,
 \eeq 
which says that the normalized average of the 
meter observable $B$ approximates the average of $A$. 

This is easily done by choosing $H$ of the form
\beq
\lbl{eqN50}
H := A \tnr G \q \mbox{where $G: M \rightarrow M$.}
\eeq 
This makes
\beq
\lbl{eqN55}
r(\epsilon) := e^{-i\epsilon H} (s \tnr m) = 
s \tnr m -i \epsilon As \tnr Gm + O(\epsilon^2),
\eeq
so 
\beq
\lbl{eqN60} 
\lim_{\epsilon \goesto 0} 
\frac{\lb r(\epsilon), 
(I \tnr B) r(\epsilon) \rb }  
{\epsilon} 
= -i \lb m, (BG - GB) m \rb \lb s , As \rb, 
\eeq 
and we need only choose $m, B,$ and G such that 
\beq
\lbl{eqN70}
1 =  -i \lb m, (BG - GB) m \rb  = 2 \Im \lb m, BGm \rb
\q, 
\eeq 
where $\Im$ denotes imaginary part.
Later we shall show how this can be done using a two dimensional meter space.  
Assuming that, the normalized average meter reading \re{eqN30} will 
approximate $\lb s, As \rb$ for small $\epsilon$.  

In addition, equation \re{eqN55} suggests that
the post-measurement state of 
$S$ should differ only slightly from $s$ for small $\epsilon$. 
(This will be examined below; for the moment we assume it.) 
That will be what we mean by a ``weak'' measurement of $\lb s, As \rb$.
For future reference, we formalize the discussion with the following 
definition. 
\begin{definition}
\lbl{def1}
Let $S$ be a quantum system and  $A$ an observable on $S$.
Let $s$ be a state of $S$ of which we can make an arbitrary number 
of copies. 
A {\em weak measurement protocol} is a procedure which can determine the
expectation $\lb s, As \rb$ of $A$ in the state $s$ to arbitrary accuracy
while perturbing each copy of $s$ used in the procedure 
by an arbitrarily small amount.  
\end{definition} 
Assuming that we can attain \re{eqN70}, and assuming the weakness of the
measurement, the above discussion describes a ``weak measurement protocol''
in the sense of the definition.  

``Weakness'' of the measurement means that the state of $S$ corresponding
to the state of $S \tnr M$ after measurement of $I \tnr B$ 
is close to  $s $ for small
$\epsilon$.%
\footnote{Since everything in sight is finite dimensional and all norms
are equivalent in finite dimensions, ``close'' may be interpreted relative to
any convenient norm.
} 
To see this, first recall that  
measurement of $I \tnr B$ will change the state 
$r(\epsilon) := e^{-i\epsilon H} (s \tnr m) $ by projecting it 
onto one of the eigenspaces of $I \tnr B$.  Each such eigenspace is the
range of a projector $I \tnr P_Q$, where $Q$ is an eigenspace of $B$.
For small $\epsilon$ 
the state $r(\epsilon)$ is close to $s \tnr m$,
so $(I \tnr P_Q)r(\epsilon)$ is close to 
$(I \tnr P_Q) (s \tnr m) = s \tnr P_Q m$.  The state of $S$ (expressed
as a density matrix) is then close to $\trM  P_{s \tnr P_ Q m} = P_s$, which
corresponds to the vector state $s$.
%
%\footnote{
%In the literature of ``weak measurements'', proofs of ``weakness'' 
%are generally omitted even in infinite dimensions, where the very 
%definition of ``weakness'' is not obvious, much less the proof.
%}
%
\section{``Weak values'' of a quantum observable}
\label{sec: weakv}
The last section defined ``weak measurement protocol'' 
and outlined an example of such, leaving details of its verification to later.
Next we discuss how such weak measurement protocols are typically used in
the literature to obtain ``weak values'' of a quantum observable.  

Before starting, we warn the reader that we think that typical 
characterizations of ``weak values'' involve an implicit logical fallacy.  
To expose
this fallacy, we shall need to present this characterization as it is typically
done in the literature (though usually less explicitly), and therefore
will sometimes need to temporarily use language which is logically questionable.  
When using such language, we shall point out that it is questionable.

We continue the discussion of the weak measurement protocol
given in the previous section.  We saw that for a system $S \tnr M$
in state $r(\epsilon) := e^{-i\epsilon H}(s \tnr m)$ 
(with $H$ as specified there), 
the normalized average value of $I \tnr B$, 
$\lb r(\epsilon), (I \tnr B) r(\epsilon) \rb/\epsilon $, 
approximates the average
value $\lb s, As \rb$ of $A$ in state $s$  for small $\epsilon$.  
Moreover, for small $\epsilon$, after the measurement of $I \tnr B$  
in state $r(\epsilon)$, the state of $S$ is close to its initial state $s$. 

Suppose that immediately after measuring $I \tnr B$, we 
``postselect''  to a given final state $f$.  
Mathematically, postselection corresponds to measuring $P_f \tnr I$ in 
$S \tnr M$.  
The physical picture is that we are asking the question ``Is $S$ in state 
$f$?''.   If the answer is ``yes'', we say that the postselection was 
successful and record the previous measurement of
$I \tnr B$, if ``no'' (in which case $S$ is in a mixture of pure states 
orthogonal to $f$), we discard the previous measurement.  

Averaging the recorded measurements 
gives the conditional expectation of $I \tnr B$ in state $r(\epsilon)$ 
given successful postselection to $f$, 
which we shall denote by $E_\epsilon (B|f)$:
\beq
\lbl{eqN85}
E_\epsilon(B|f) := 
\frac{\lb r(\epsilon), (P_f\tnr B) r(\epsilon) \rb}
{\lb r(\epsilon), (P_f \tnr I) r(\epsilon) \rb} 
\q.
\eeq 
Here we are modeling measuring $B$ then immediately postselecting 
as measuring $P_f \tnr B = (P_f \tnr I)(I \tnr B)$.  
(We could also think of the measurement
of $B$ and postselection as occurring simultaneously; since $P_f \tnr I$
and $I \tnr B$ commute, we can measure them simultaneously.)

The conditional expectation $E_\epsilon (B|f)$ is $O(\epsilon)$. 
It is tempting to normalize by dividing 
by $\epsilon$ and take a limit as $\epsilon \goesto 0$, 
resulting in a normalized conditional expectation
which we denote by $\NE (B|f)$,
\beq
\lbl{eqN87}
\NE(B|f) := \lim_{\epsilon \goesto 0} \frac{E_{\epsilon}(B|f)}{\epsilon}
\q,
\eeq 
and then identify $\NE(B|f)$ with the 
conditional expectation of $A$ in state $s$ given successful postselection.  
(We think this identification fallacious, for reasons given below.)

The normalized conditional expectation $\NE(B|f)$ is called a 
``{\bf weak value}'' of $A$.%
\footnote{The term ``weak value'' is 
rarely precisely defined in the literature.
Some authors 
introduce it in ways which may seem different but 
are conceptually almost identical.  
A full discussion of how to recast the formulation of AAV into the terms
above is given in \cite{parrott}. 
} 
Note, however, that this terminology may be potentially misleading
because the ``weak value of $A$'' might depend not only on $A$ (and other
data associated with the original system $S$ of interest such as 
$s, f \in S$), but also on data external to $S$ such as the meter 
observable $B$ and the meter state $m$.  

Most of 
the traditional literature calculates this ``weak value'' as%
\footnote{AAV \cite{AAV} 
calulates it as $\lb f, As \rb / \lb f, s \rb$, a quantity
which need not even be real, though a footnote in that paper suggests that
the authors realize that the real part should be taken.  
Though subsequent literature sometimes quotes the AAV formula, the real
part seems to be always taken in actually applying it. Ref.\ \cite{parrott}
explicitly identifies a possible error in the mathematics of AAV which 
could result in the omission of the real part.  } 
\beq
\lbl{eqN100} 
\Re \frac{\lb f , As \rb}{\lb f, s \rb} \q,
\eeq
where $\Re$ denotes real part.  It is usually called  something
like ``the weak value
of $A$ in initial state $s$ and postselected state $f$,''
with all references to data external to $S$ suppressed.
This risks leaving the impression that ``weak values'' measure
something intrinsic to $S$, independently of details of the meter system. 

Using language which ignores the meter system may seem 
superficially reasonable because the normalized expectation
of $I \tnr B$ in state $r(\epsilon)$  unquestionably approximates 
the expection $\lb s , As \rb$ of $A$ in state $s$, as shown in
equation \re{eqN40}, independently of details of the meter system.  
This is as it should be for a good meter.  

For example, all voltmeters which draw sufficiently small current
(the analog of sufficiently weak coupling) are expected to read the
about the same voltage.  
If meter A read ``voltage'' 50 and meter B read ``voltage'' 700
and both disturbed the system being measured negligibly (i.e., drew about 
the same small current), one would conclude that the two meters were measuring  
different things, of which at least one and possibly both 
should not be called ``voltage''.

Postselection replaces the (normalized) expectation
\re{eqN40} of the meter observable (which is also the expectation
$\lb s, As \rb$ of $A$) with the (normalized) conditional expectation
$\NE (B|f)$ (which is not obviously identifiable with anything intrinsic
to the system $S$ such as a conditional
expectation of $A$).  This makes it questionable to suppress details of the
meter from the language.

We think that 
a transition from unconditional to conditional expectation of $A$ would
involve a 
logical fallacy which we now discuss in detail.%
\footnote{Whether the ``weak value'' literature actually assumes this
fallacy may be controversial.  The only paper known to me which clearly 
does not is \cite{jozsa}.  Nearly all papers on ``weak values'' 
do seem to me to suggest some identification of the traditional
weak value \re{eqN100} with an intrinsic property of $A$ (and $f$ and $s$), 
though many are so vaguely written that it can be hard to say precisely 
what they might have intended.} 

Consider the two sentences:
\begin{enumerate}
\item
  The average value of the normalized (i.e., by division by $\epsilon$) 
meter measurement equals 
the average value of $A$ (in the limit $\epsilon \goesto 0$). 
\item
The average value of the normalized meter measurement conditioned 
on postselection to $f$ equals 
the average value of $A$ conditioned on postselection to $f$
(in the limit $\epsilon \goesto 0$).
\end{enumerate}
Here ``meter measurement'' refers to measurement of $I \tnr B$.

We have seen that Sentence 1 is true. But Sentence 2 is either false, 
meaningless, or 
tautological, depending on how it is interpreted.

In order to speak meaningfully of 
``the average value of $A$ conditioned on postselection
to $f$'', 
we need to say how this quantity is measured.  If we measure
$A$ in $S$, successfully postselect to $f$, and average the results, 
we do {\em not} necessarily
obtain the traditional ``weak value \re{eqN100}, 
$\Re (\lb f, As \rb / \lb f, s \rb)$ (which is obtained from a specific
choice of $B$ and $G$).  
This is a simple
calculation which will be done below. 

Moreover, we shall show that many {\em different}
``weak values'' 
can be obtained with other choices of $B$ and  Hamiltonian $H := A \tnr G$ 
(i.e., weak values  
are not unique).
This shows that independently of the correctness of 
calculations of the traditional ``weak
value'' \re{eqN100}, sentence 2 cannot be true 
if  
``the average value of $A$ conditioned on postselection to $f$''
refers to normal measurements  
in $S$ (without reference to the meter system $M$).  

To what could it refer?  
If it refers to measurement in $M$ of the normalized 
average value
of $B$ postselected to $f$ (in the limit $\epsilon \goesto 0$), 
then sentence 2 becomes a tautology, true
by definition and containing no useful information.   

We now perform the simple calculation of
the average value of $A$ conditioned on postselection to $f$,
assuming that we are referring to normal measurements in $S$ (forgetting for
the moment about the meter system $M$).
For notational simplicity, we assume that 
$A$ has just two distinct eigenvalues $\alpha_1 \neq \alpha_2$ with
corresponding normalized eigenvectors $a_1, a_2$. 

After $A$ is measured, $S$ is in state $a_i$ with probability 
$|\lb a_i, s \rb |^2$, $i = 1,2$. Subsequently, the postselection succeeds with
(conditional) probability $| \lb f, a_i \rb |^2$. 
The total probability that the postselection succeeds is 
$$
\sum_{i=1}^2 |\lb a_i, s \rb |^2 | \lb f, a_i \rb |^2
\q.
$$
Hence the conditional expectation of $A$ given that the postselection
succeeds is
\beq
\lbl{eqN110}
\frac
{\alpha_1 |\lb a_1, s \rb |^2 |\lb f, a_1 \rb |^2  
+ \alpha_2 |\lb a_2, s \rb |^2 | \lb f, a_2 \rb |^2 
}
{  |\lb a_1, s \rb |^2 | \lb f, a_1 \rb|^2  
+  |\lb a_2, s \rb |^2 | \lb f, a_2 \rb|^2
}
\q.
\eeq
It is easy to construct examples in which this does not equal 
the traditional ``weak value'' \re{eqN100}, 
$\Re( \lb f, As \rb / \lb f, s \rb$.  
For example, \re{eqN110} is a convex
linear combination of $\alpha_1, \alpha_2$, whereas it is well known that
\re{eqN100} need not be.  
This is emphasized by the provocative title of AAV \cite{AAV}:
``How the result of a 
measurement of a component of the spin of a spin-1/2 particle can
turn out to be 100''.
(When \re{eqN100} is not a convex linear combination
of the eigenvalues of $A$, it is called a ``strange'' weak value.) 
\section{Weak values are not unique} 
\label{sec: nonunique}
The last section exposed what we believe to be a logical fallacy implicit in 
the very concept of ``weak value''.  Perhaps not  
all will accept this, but even 
those who do may wonder if the traditional ``weak value'' 
\re{eqN100}, $\Re(\lb f, As \rb / \lb f, s \rb )$, expresses some intrinsic
property of the system $S$.  The traditional literature would encourage such 
a belief because the only ``weak values'' which appear, to our knowledge, 
are  \re{eqN100}  
with or without the real part, and experiments verify only \re{eqN100}.  

If other ``weak values'' can be obtained by similar reasoning, then that would
cast doubt on any belief that \re{eqN100} expresses some intrinsic property
of observable $A$ in system $S$.
This section continues the reasoning of the ``weak measurements'' section
to calculate the traditional ``weak value'' \re{eqN100} and then calculates
other ``weak values'' by similar reasoning.  Notation is as defined there. 
Recall in particular that
\beq
\lbl{eqN200b1} 
r(\epsilon) := e^{-i\epsilon H (s \tnr m)} = e^{-i A \tnr G}(s \tnr m) 
= s \tnr m -i\epsilon As \tnr Gm + O(\epsilon^2)  
\eeq
with $H := A \tnr G$ and both $G$ and 
the meter observable $B$ yet to be defined. 

We shall calculate the normalized conditional expectation $\NE(B|f)$ 
of the meter reading given
postselection to $f$ defined in equations \re{eqN85} and \re{eqN87}: 
by:
\beq
\lbl{eqN200}
\NE(B | f) := \lim_{\epsilon \goesto 0} 
\frac{1}{\epsilon}
 \frac{
\lb r(\epsilon), ( P_f \tnr B) r(\epsilon) \rb }
{\lb r(\epsilon) , (P_f \tnr I) r(\epsilon)\rb}
.
\eeq 

First we calculate, recalling assumption \re{eqN20} that $\lb m, Bm \rb=0,$ 
\begin{eqnarray}
\lbl{eqN210}
\lefteqn{ 
\frac{\lb r(\epsilon), (P_f \tnr B) r(\epsilon) \rb}
{\epsilon}
}
&& \nonumber\\
&=&
\frac{1}{\epsilon}
\left[ \lb s, P_f s \rb \lb m,  Bm \rb + \lb s \tnr m , -i \epsilon
P_f As \tnr BGm \rb \right. \nonumber\\
&& \q \left. +\  \lb -i \epsilon As \tnr Gm , P_f s \tnr  Bm \rb +
O(\epsilon^2) \right] 
\nonumber\\
&=& 
  \lb s, -i P_f As \rb \lb m, BGm\rb + \lb -i As, P_f s\rb\lb Gm, Bm\rb
+ O(\epsilon)
 \nonumber\\
&=& -i[\lb f, As\rb\lb s, f \rb \lb m, BGm\rb - \lb f, s \rb \lb As, f \rb
\lb Gm, Bm \rb ] + O(\epsilon)
\nonumber\\
&=& -i [ \lb f, As \rb \lb s, f \rb \lb m, BGm \rb - \lb As, f \rb
\lb f, s \rb \lb BGm, m \rb ] + O(\epsilon) 
.
\end{eqnarray}
Next note that
$$
\lim_{\epsilon \goesto 0} \lb r(\epsilon), (P_f \tnr I)r(\epsilon) \rb
= \lb s \tnr m, (P_f \tnr I)(s \tnr m) \rb = \lb s, P_f s \rb = 
\lb f, s\rb \lb s, f \rb.
$$
Combining this with \re{eqN210} gives
\begin{eqnarray}
\lbl{eqN220}
\NE(B | f) &=& 
\frac{ -i [ \lb f, As \rb \lb s, f \rb \lb m, BGm \rb - \lb As, f \rb
\lb f, s \rb \lb BGm, m \rb ] 
}
{\lb f, s \rb \lb s, f \rb}
\nonumber\\
&=& -i\left[ \frac{ \lb f, As \rb}{\lb f, s \rb} \lb m, BG m\rb -  
\frac{\lb As, f \rb}{\lb s, f \rb}\lb BGm, m \rb \right]
\nonumber \\
&=& 
2 \Im  \frac{\lb f, As \rb \lb m, BGm \rb}{\lb f, s \rb}
\q,
\end{eqnarray}
when $\lb f, s \rb \neq 0$ and is undefined if $\lb f, s \rb = 0.$ 
The following calculations assume $\lb f , s \rb \neq 0$.

We still need to specify $M$ and $m$ and $B$ and $G$.  
We have been operating under
assumption \re{eqN20} that $\lb m, Bm \rb = 0.$  
To this we must adjoin condition \re{eqN70}, namely 
\beq
\lbl{eqN230}
\Im \lb m, BGm \rb = \frac{1}{2}
\q,
\eeq
to assure that the normalized average measurement of $I \tnr B$ be 
$\lb s, As \rb$.
Write 
\beq
\lbl{eq 860}
\lb m, BGm \rb = \rho + \frac{1}{2} i  \q \mbox{with $\rho$ real.}
\eeq 
Then \re{eqN220} becomes 
\beq 
\lbl{eqN240}
\NE(B|f) = 
2 \Im  \frac{\lb f, As \rb \lb m, BGm \rb}{\lb f, s \rb}
=
\Re \frac{\lb f, As \rb}{\lb f, s \rb} 
+ 2 \rho \Im \frac{\lb f, As \rb}{\lb f, s \rb}
\q.
\eeq
The first term in \re{eqN240} is the ``usual'' weak value
$\Re( \lb f, As \rb / \lb f, s \rb)$, but we shall show that 
when $\Im (\lb f, As \rb / \lb f, s \rb) \neq 0,$ 
the second term can be chosen
arbitrarily by adjusting the value of $\rho = \Re \lb m, BGm \rb$.    

To see that any number can be obtained for $\Re \lb m, BGm \rb$ with 
$\lb m, Bm \rb = 0$ and   
$\Im \lb m, BGm \rb = 1/2$, take the meter space $M$ to be two-dimensional
with orthonormal basis $m, m^\perp$, and define $G$ and $B$ by the
following matrices with respect
to this basis:
\beq
\lbl{eqN250}
G := 
\left[
\begin{array}{ll}
0 & 1  \\
1  & 0
\end{array}
\right]
 \q \mbox{and}\q 
B := 
\left[
\begin{array}{ll}
0 & \rho + i/2\\
\rho - i/2 & 0
\end{array}
\right] 
\eeq
with $\rho$ real.
Then $BG$ has the following form, where entries denoted ``$*$'' have not
been calculated because they are irrelevant to calculation of 
$\lb m , BGm \rb$ (which is the upper left entry of BG):
\beq
\lbl{eqN260}
BG = 
\left[ 
\begin{array}{rr}
\rho + i/2 & * \\
** & * \\
\end{array} 
\right]
= 
\left[ 
\begin{array}{rr}
\lb m, BGm \rb & * \\
** & * 
\end{array}
\right].
\eeq
This shows that when $\Im (\lb f, As \rb / \lb f, s \rb) \neq 0$, 
by varying $\rho$, one can obtain any number whatever as a ``weak value''
for $A$: weak values are not unique.
 
If it happens that $ \Im (\lb f, As \rb / \lb f, s \rb) = 0$, 
then the traditional ``weak value'' $ \Re (\lb f, As \rb / \lb f, s \rb) $ 
is uniquely obtained {\em by the above method}.  However, other methods
can yield other ``weak values''.  An example is given in \cite{parrott}.  
\section{Obtaining non-traditional ``weak values'' in the context of AAV}
\label{sec: AAV}
Von Neumann \cite{vN} and AAV \cite{AAV} both used as a meter space
the infinite dimensional space $L^2(\bR)$ 
of all complex-valued, square-integrable
functions on the real line, known to physicists as the Hilbert space of a 
single spinless particle in one dimension.  They both used a preparation
Hamiltonian of the form $H := A \tnr P$, where $P$ is 
the usual momentum operator 
defined by 
$(Pf)(q) := -i\, df/dq$, 
and their meter operator (which 
we called $B$) is the usual position operator $Q$ defined by  
$Qf(q) := q f(q)$.  
AAV used an $\epsilon$-dependent Gaussian as the square $m^2$ of their 
real meter
function $m$ (so that initially the meter position had a Gaussian
distribution with a variance depending on a parameter $\epsilon$).  
Thus their setup is algebraically quite similar to ours
(and of course ours was inspired by theirs).  There are some apparent 
differences which are discussed in detail in \cite{parrott}, but these turn
out to be inessential.%  
\footnote{We mention two here to forestall confusion for readers who
may be consulting AAV.  They use a preparation Hamiltonian $ - A \tnr Q$
instead of our $ A \tnr P$, and their meter observable is $I \tnr P$
instead of our $I \tnr Q$.  Since the Fourier transform isomorphism 
takes $Q$ to $-P$ and  $P$ to $Q$, 
this difference is immaterial.
\s
Also, they use  
an $\epsilon$-dependent meter state $m[\epsilon)](\cdot) $ 
which in our notation would be 
$m[\epsilon](q) := m(q \epsilon)\sqrt{\epsilon}$
in conjunction with an $\epsilon$-independent preparation operator
which in our notation would be
$e^{i A \tnr P}$.  This makes $\lb e^{i A\tnr P} (s \tnr m[\epsilon]) , 
(P_f \tnr Q)e^{iA \tnr P}(s \tnr m[\epsilon]) \rb$ 
of constant order (i.e., $O(1)$ 
instead of $O(\epsilon)$ as in our formulation), so that our normalization 
by dividing by $\epsilon$ is unnecessary.  The details are worked 
out in \cite{parrott}, resulting in the conclusion that their setup is 
algebraically equivalent to ours.  
} 

The calculations done above 
are rigorous for finite-dimensional $S$ and $M$, and 
still algebraically correct in infinite dimensions.  In the physics literature,
such algebraic calculations are typically accepted as ``proofs''.
If we relax mathematical rigor to this extent, we obtain from the above
a very simple ``proof'' in the AAV context 
of the ``usual'' weak value $\Re (\lb f, As \rb / 
\lb f, s \rb)$ by taking 
\beq
\lbl{eqN270}
M := L^2(\bR), \q  G := P,\q  B := Q , \q m(q) := 
\left[\frac{1}{\sqrt{2\pi}}
e^{- q^2/2}\right]^{1/2} .
\eeq

This ``proof'' is deeply flawed because the starting
equation \re{eqN200b1},
$$
e^{-i\epsilon (A \tnr G)}(s \tnr m) 
= s \tnr m - i \epsilon As \tnr Gm + O(\epsilon^2)
,
$$ 
would probably be difficult to justify for our unbounded $G := P$. 
All of the ``proofs'' of the ``usual'' weak value formula 
which we have seen in the literature
rely on uncontrolled approximations like this.  In honesty, they should be
called something like ``algebraic motivations'' instead of proofs. 

If we {\em are} willing to accept uncritically such uncontrolled approximations,
we can obtain arbitrary weak values in an AAV-type framework by 
taking $ B$ and $m$ as in equation \re{eqN270} and 
\beq
\lbl{eqN275}
G : P + \rho Q \q \mbox{ with $\rho \neq 0$ real.} 
\eeq
This results in 
\beq
\lbl{eqN280}
\lb m , BG m \rb = \rho  + \frac{i}{2} \q,
\eeq
so that when $\Im ( \lb f, As \rb / \lb f , s \rb) \neq 0$, any ``weak value''
whatever can be obtained using the preparation Hamiltonian 
$H :=  A \tnr (P + \rho Q)$.   

%Since $\delta$ can be arbitrarily small, anyone who claims that the ``usual''
%weak value formula gives the only experimentally possible result 
%should be obligated to explain how the von Neumann/AAV Hamiltonian
%$H :=  A \tnr P$ can be guaranteed in any experimental
%situation, and assuming that, how the Hamiltonian $A \tnr P$ can be 
%experimentally distinguished from $A \tnr (P + \delta Q)$ for arbitrarily 
%small $\delta$. On the level of uncontrolled approximations, this will 
%probably be impossible.  

It is almost immediate that  
\beq
\lbl{eqN285}
e^{-i Q^2 \rho/2} P e^{i Q^2 \rho/2} = P + \rho Q 
\eeq
because for any $g \in L^2 (\bR)$, $(e^{i Q^2 \rho/2}g)(q)
= e^{i q^2 \rho/2}g(q)$ and $P := -i\, d/dq$.%
\footnote{We say ``almost'' because
a rigorous verification would
require careful specification of the domain of $P$, which
we have not discussed.
}
In other words, our preparation Hamiltonian 
$P + \rho Q$ is formally (i.e., algebraically, ignoring analytical
subtleties)  unitarily equivalent to AAV's.  This unitary equivalence
carries the Gaussian  meter state $m(q)$ (which is effectively that
used by AAV) into 
\beq
\lbl{eqN290} 
q \mapsto e^{-iq^2\rho/2} m(q)\q, 
\eeq
which still defines a Gaussian probability
distribution on position space.

A rigorous derivation of the ``usual'' weak value in the AAV framework
(i.e. meter space $L^2(\bR)$ and using the AAV Hamiltonian and meter state
was given in \cite{parrott}), along with an extension of that argument
which yields   
a rigorous proof that weak values are not
unique even using the AAV Hamiltonian $A \tnr P$ 
(but with the slightly different meter state \re{eqN290}).  
\section{Remarks}
\label{sec: remarks}
\begin{enumerate}
\item
Definition 1 of ``weak measurement protocol'' assumed a ``reproducible'' 
state $s$ of $S$, i.e., a state 
for which  an arbitrary number of copies are available.%
\footnote{This does not contradict the no-cloning theorem because is is not
required that an arbitrary state can be copied, only that we have a device
which can make any number of copies of some particular state in
which we are interested.}
All weak measurement schemes known to us require this assumption. 
This is because for very weak coupling between the system $S$ of interest
and the meter system, a large number of meter measurements may be needed 
to obtain a reliable average.
\s
In finite dimensions,
a reproducible state may be considered ``known'' in the sense that 
its components with respect to a given basis can be estimated to arbitrary
accuracy by quantum tomography (\cite{N/C}, pp. 389ff).  But if $s$ is 
known, then so is $\lb s, As \rb$, and one wonders what is the point 
of finding a weak measurement protocol to measure $\lb s, As \rb$. 
It is true that such a protocol can perform the measurement with negligible
effect on the copies of $s$ used in the measurement, but it's hard to see 
how this could be useful when an arbitrary number of copies of $s$ were 
available from the start.  
\item
Once one realizes that it is either tautological or incorrect to identify
the normalized conditional expectation $\NE(B|f)$ of $I \tnr B$ 
with the 
``expectation of $A$ in state $s$ conditional on postselection to $f$'',
the concept of ``weak measurement'' seems to collapse.  What is left? 
\s
When the denominator $\lb f, s \rb$
of the traditional weak value $\Re(\lb f, As\rb /
\lb f, s \rb)$ is small, the weak value can be very large relative to 
the norm of $A$.  
This  is sometimes considered as an ``amplification'' effect.
Assuming that the experimental procedure faithfully implements
the mathematics leading to that weak value, 
the ``amplification'' is real in the sense 
that the normalized conditional (i.e., on successful postselection) 
expectation
of the meter can greatly exceed all eigenvalues of $A$.  
Reference  \cite{Kwiat} reports such amplification.
(For an expository account, see \cite{Resch}.)
However, in a general context, it seems unclear precisely what is being 
``amplified'', if ``amplification'' is  considered a linear
process as usual.
It cannot be $\lb s, As \rb$ because   
the traditional weak value $\Re(\lb f, As \rb / \lb f, s \rb )$ 
is not linear in $\lb s, As \rb$.  
Indeed, one could have $\lb s, As \rb = 0$ with the 
weak value $\Re(\lb f, As \rb / \lb f, s \rb)$ nonzero.  
\item
Nearly all of the ``weak value'' literature (with the notable exception
of a recent paper \cite{jozsa} of Jozsa) presents the traditional
``weak value'' $\Re (\lb f, As \rb / \lb f, s \rb )$ as if it were 
the only theoretical possibility, and as if it were experimentally inevitable.
Even the Jozsa paper does not comment on the implications of the nonuniqueness
of weak values. 
\s
In view of the nonuniqueness of weak values, claims of the experimental
inevitability of the traditional weak value should be carefully scrutinized.
It seems strange that papers describing complicated 
experiments to measure weak values generally ignore this crucial point. 
It should not be taken for granted that the AAV Hamiltonian and meter state
can be assumed without detailed justification.  
\end{enumerate}
\section{Summary}
\label{sec: summary}

We have defined ``weak measurement'' and ``weak value'' 
of a quantum observable, and have given a rigorous proof in 
a finite dimensional context that weak values of a quantum observable $A$
in a state $s$ of a quantum system $S$ 
are not unique.   
This implies that weak values
need not be  intrinsic to the system $S$ being ``weakly'' measured: 
in general they may depend also on details of the meter system (such
as the meter state).  

Most of the ``weak value'' literature presents the traditional weak value
$\Re ( \lb f, As \rb / \lb f , s \rb)$ as if it were experimentally 
inevitable, without mentioning the possibility of other ``weak values''.
We suspect that this may arise from the logical fallacy of identifying
the (normalized) expectation of a ``meter observable'' 
conditional on postselection 
to $f \in S$ with the expectation of $A$ conditional on postselection
to $f$.  The (normalized) expectation of the meter observable 
does equal the expectation
of $A$, but there is no reason that the conditional expectations 
should be equal. 

Traditional ``weak values'' are usually associated with a particular
infinite dimensional context introduced in AAV \cite{AAV}. 
Though the analysis given in this paper is rigorous only for finite
dimensional systems, it can be rigorously extended to the AAV context,
yielding multiple weak values in that context;
details are given in \cite{parrott}.  

The nonuniqueness of ``weak values'' 
suggests that any claims of the inevitability
of the ``traditional'' weak value $\Re(\lb f, As \rb / \lb f, s \rb)$ 
in experimental weak measurements 
will probably have to
be based on arguments for the universality of something close to
the precise AAV setup
(e.g., Hamiltonian, meter observable, and meter state).
Since there seems to exist no argument in the literature that this precise
setup can be realized in {\em any} experimental situation, much less be
inevitable in all,
such an argument would probably have to break new ground. 
\section{Afterword} 
I thank an anonymous referee for drawing my attention to an interesting paper 
of R.\ Jozsa \cite{jozsa} which motivates 
(formally, using uncontrolled approximations)  
a ``Theorem'' relating 
weak values to
the real and imaginary parts of the complex AAV
weak value $\lb f, As \rb/ \lb f, s \rb$, along with the mean and variance 
of the meter observable, assuming  that 
the meter state satisfies a Schroedinger
equation before the preparation Hamiltonian is applied.  

In our notation (and under our assumption that $\lb m, Qm \rb = 0$)
this Theorem states that (recall that $NE(Q|f)$ is the normalized 
expectation of the meter observable conditional on postselection
to $f \in S$)
\beq
\lbl{eq1000}
\NE(Q|f) = \Re 
\frac{\lb f, As \rb}{\lb f, s \rb}
+ \Im  \frac{\lb f, As \rb}{\lb f, s \rb}
k \left. \frac{d \lb m, Q^2 m \rb}{dt}\right|_{t=0^-} \q, 
\eeq
where $Q$ is the meter observable of our Section 5 
(i.e., $Qf(q):= qf(q)$ for $f \in L^2(\bR))$ 
and $k$ is the mass of the meter pointer.%
\footnote{
The reader should be warned that the proof of the stated ``Theorem'' 
giving this 
relation seems to require an unstated hypothesis that  
a boundary term in a partial integration (his equation (14))
can be dropped; otherwise the coefficient of 
$\Im (\lb f, As\rb/\lb f, s \rb)$
in
\re{eq1000} will not necessarily be as stated. 
} 
(Because the meter state $m$
 is assumed to satisfy a Schroedinger equation, both it and 
$\lb m, Qm \rb$ are time-dependent.) 

If we assume that the coefficient of  
$ \Im  (\lb f, As \rb / \lb f, s \rb )$ can be arbitrary (as seems 
reasonable and probably provable under mild additional hypotheses),
this is  like our \re{eqN240} for this particular setup (i.e., meter state
assumed to satisfy a Schroedinger equation).  In particular, this 
motivates the existence of arbitrary weak values for a given observable
$A$.  However, \cite{jozsa} does not comment on the implications of the
fact that weak values are not unique.     

I also thank that referee for pointing out a potential 
ambiguity in the original exposition, 
which has been corrected in this version.  

Another referee objected to the paper's suggestion that ``we think that
typical characterizations of `weak values' involve an implicit logical 
fallacy''. 
Since this is a personal opinion which is clearly identified as such, 
I saw no reason to excise it from this later version.  

Most presentations
of weak values in the literature are so vaguely written that is is difficult
to discern precisely what the authors might have meant.  
I think it would be very difficult
to explicitly and convincingly motivate the concept of ``weak value'' without
comitting some such logical fallacy.  The exposition required some such
motivation, so I made my best guess, and it still seems to me the best guess. 
If some reader thinks that guess obviously wrong, he or she is invited to
provide an alternate motivation which is clear, correct, and consistent with the
traditional literature.

More information about the submission history of this paper along with all 
referees' reports and comments on them
can be found on my website,
www.math.umb.edu/$\sim$sp in the ``papers'' page.  
%

%MARK

\end{document}